\documentstyle[aps,prd,epsf]{revtex}
\def\bm#1{\mbox{\boldmath{$#1$}}}
\begin{document}
\title{
Large Scale Features of Rotating Forced Turbulence }
\author{ Jos\'e Gaite$^{+}$,
David Hochberg$^{++}$, and Carmen Molina-Par\'\i s$^{+++}$ }
\address{
$^{+}$Instituto de Matem{\'a}ticas y F{\'\i}sica Fundamental,
CSIC, Serrano 123, 28006 Madrid, Spain
\\
$^{++}$Centro de Astrobiolog\'\i{}a (CSIC-INTA), Carretera de
Ajalvir Km. 4, 28850 Torrej\'on de Ardoz, Madrid, Spain
\\
$^{+++}$Department of Applied Mathematics, University of Leeds,
Leeds LS2 9JT,United Kingdom}

\maketitle

\begin{abstract}

Large scale features of a randomly isotropically forced incompressible
and unbounded rotating fluid are examined in perturbation theory. At
first order in both the random force amplitude and the angular
velocity we find two types of modifications to the fluid equation of
motion.  The first correction transforms the molecular shear viscosity
into a (rotation independent) effective viscosity. The second
perturbative correction leads to a new large scale non-dissipative
force proportional to the fluid angular velocity in the slow rotation
regime. This effective force does no net work and alters the
dispersion relation of inertial waves propagating in the fluid. Both
dynamically generated corrections can be identified with certain
components of the most general axisymmetric ``viscosity tensor'' for a
Newtonian fluid.

\end{abstract}


\section{Introduction}
\label{sec:intro}

The special features of turbulence in the presence of rotation have
attracted the interest of many
authors~\cite{Greenspan,Cambon1,Jacquin,Dubrulle}.  Relying on some
experiments, the methods of study used have ranged from analytic
approaches to numerical
simulations~\cite{Cambon1,Veera,Cambon2,Godeferd}.  The central theme
in rotating fluids is the effect of the Coriolis force, which induces
anisotropy (there is a preferred direction, that of the rotation
axis). This anisotropy is extreme in the limit of fast rotation, which
actually forces the flow to become two-dimensional (Proudman-Taylor
theorem)~\cite{Greenspan}. In this work we apply perturbation theory
to the randomly forced Navier-Stokes equation with Coriolis force as a
model for the turbulent regime of a rotating fluid. The perturbative
study of the ordinary randomly forced Navier-Stokes equation, in
combination with the renormalization group (as an improvement of
perturbation theory), has a long tradition~\cite{FNS,YO,McComb}.

The addition of the Coriolis force, induced by the rotation of the
fluid, introduces one additional parameter, the angular velocity
$\Omega$ or, in dimensionless form, either the Rossby or Ekman numbers
(in addition to the Reynolds number)~\cite{Greenspan}. Let us focus on
the Ekman number, $Ek = \nu/(\Omega L^2)$, that depends on the
viscosity $\nu$ and a scale $L$, roughly associated with the size of
the fluid system. The Ekman number gives the relative importance of
the viscosity and Coriolis forces. We will {\it assume} henceforth
that for small $\Omega$ ($< {\nu/L^2}$) the turbulence is isotropic
and the only relevant parameter is the viscosity. In this limit, the
results of the study of the ordinary randomly forced Navier-Stokes
equation hold (the random force is always assumed isotropic).

For larger $\Omega$ we will encounter new features. In fact, the only
restriction on perturbative correction terms is that they respect the
basic {\it symmetry} of the equations, in our case, the axial symmetry
about the rotation axis. We will see that perturbation theory
generates new terms fulfilling these symmetry constraints. Therefore,
one must find the complete set of allowed terms that can arise in
perturbation theory. We will determine all the terms that can be
represented by the components of an axisymmetric ``viscosity tensor''
and, in particular, the ones that arise at first order in perturbation
theory. The part of this ``viscosity tensor'' that is
pair-antisymmetric in the indices plays a significant role; however,
it does not lead to dissipation, and therefore, is not truly viscous.

As in homogeneous and isotropic turbulence, we assume that the
physical region of study is sufficiently far from the surfaces, where
the boundary conditions are imposed, for them not to have any {\it
direct} effect, except the presence of the scale $L$. In contrast to
ordinary turbulence, this condition only implies that we can have
homogeneous turbulence but, due to rotation, it cannot be
isotropic. It is pertinent to mention here that the possibility of
anisotropic forced turbulence, and precisely with axial symmetry, has
already been considered~\cite{Chek}. In this reference, however, the
authors assume that the breakdown of isotropy occurs through a random
force whose two-point correlation function depends on the anisotropy
vector $\vec{n}$. They derive a renormalized force proportional to
second and fourth powers of $\vec{n}$. In our case, we will see that
the first perturbative correction is linear in $\vec{\Omega}$, like
the Coriolis force itself.

This paper is organized as follows. In
Section~\ref{sec:basic+perturbation} we introduce the randomly forced
hydrodynamical equations with rotation.  We assume that the fluid is
incompressible and show how to formulate them as a problem of
homogeneous but anisotropic incompressible turbulence.  Fourier
analysis of the turbulent velocity field is used to organize the
perturbation expansion~\cite{McComb} in
Subsection~\ref{sec:fourier}. We introduce in
Subsection~\ref{sec:response} the linear response function. Unlike the
isotropic case, the Coriolis term leads to a {\it non-symmetric}
linear response function matrix.  In Subsection~\ref{sec:perturbative}
we define the non-linear response function and present its
perturbative expansion (slow rotation). We also compute the first
order perturbative correction to the response function, which allows
the identification of the (rotation independent) effective shear
viscosity (proportional to the cube of the Reynolds number), and a new
anisotropic force.  In Section~\ref{sec:effective-tensor} we write
down the most general axially symmetric ``viscosity tensor'', as the
existence of a preferred direction, singled out by the fluid rotation,
requires the introduction and use of axisymmetric tensors.  This rank
four ``viscosity tensor'' expresses the proportionality between the
fluid stress tensor and the rate of strain tensor. In isotropic and
homogeneous incompressible turbulence, the viscosity tensor depends
only on one parameter, the fluid shear viscosity. In the case of
rotating turbulence, and for slow rotation, we find that the
axisymmetric ``viscosity tensor'' depends on two parameters: the
molecular shear viscosity (coming from the isotropic terms of the
``viscosity tensor'') and a new one, that arises from the anisotropic
terms in the ``viscosity tensor''.  We also show that this new
parameter can be identified as the coefficient of the anisotropic
force calculated perturbatively (in the previous
Section~\ref{sec:perturbative}).  Having thus established the
equivalence between the perturbatively corrected randomly forced
Navier-Stokes equation with Coriolis force on the one hand, and a
(Newtonian) rotating incompressible fluid with an effective
axisymmetric ``viscosity tensor'' on the other, we proceed, in
Section~\ref{sec:effective}, to discuss some physical consequences of
the new terms in the perturbed fluid equations. In
Subsection~\ref{sec:axial-force} we consider the quasi-local force
induced by the anisotropic components of the ``viscosity tensor'' and
show that it is proportional to the cube of the Reynolds number, and
that it does not lead to dissipation. In Subsection~\ref{sec:waves} we
study the dynamical effects of this force on the propagation of
inertial waves. We end by discussing our results and proposing further
work on the problem of rotating turbulence.  In
Appendix~\ref{sec:app-diag} we introduce the diagrammatic
representation of the exact Navier-Stokes equation and the diagram
encoding the first order correction to the response function and in
Appendix~\ref{sec:app-perturbation} we present the technical details
needed to carry out the perturbative calculation in the slow rotation
limit.

\section{Basic equations and perturbation theory}
\label{sec:basic+perturbation}

\subsection{Equations of motion with random force in a rotating frame}
\label{sec:equations}

We start from the hydrodynamical equations for a fluid with density
field $\rho({\bf x},t)$, velocity field ${\bf u}({\bf x},t)$, pressure
$p ({\bf x},t)$, and molecular shear (tangential) and bulk kinematic
viscosities $\nu$ and $\kappa$, respectively. We assume that the fluid
is rotating with constant angular velocity ${\bf \Omega}$ along the
$\hat z$ axis and that it is subject to an isotropic random forcing
{\it per unit mass} ${\bf f}$.  The mass and momentum conservation
equations are
\begin{mathletters}
\begin{eqnarray}
\frac{\partial \rho}{\partial t} + \vec \nabla \cdot(\rho {\bf u})
&=& 0
\; ,
\label{density1}
\\
\frac{\partial {\bf u}}{\partial t} + ({\bf u }\cdot\vec \nabla)
{\bf u} &=& -\frac{1}{\rho}\vec \nabla p
+ {\nu} \nabla^2 {\bf u} +
\left[
\kappa + \nu \left(\frac{d-2}{d} \right)  \right]
{\vec \nabla}({\vec \nabla \cdot {\bf u}})
-2{\bf \Omega \wedge u}
- {\bf \Omega \wedge (\Omega \wedge x)}
+ {\bf  f}
\; ,
\label{moment1}
\end{eqnarray}
where $d$ is the number of space dimensions. The dimension of space
$d$ will be kept as a free variable, although when we consider
rotation-dependent expressions, these must be evaluated for $d=3$.
The momentum equation~(\ref{moment1}) is supplemented with a random
stirring force that leads to a statistical distribution for the
velocity field and can be used to model turbulent flows just as is
done for isotropic randomly stirred (non-rotating)
turbulence~\cite{FNS,McComb}.  Regarding the random force spectrum and
statistics, we take a Gaussian random force that is white in time,
{for simplicity}, but we allow for ({translation invariant}) spatial
correlations.  So we can write
\end{mathletters}
\begin{eqnarray}
\langle f_i (\vec x,t) \rangle =
0
\;
\; \; \; \; \; \;
{\rm and}
\; \; \; \; \; \;
\langle f_i (\vec x,t) f_j (\vec x',t') \rangle =
D_{ij}(\vec x-\vec x')
\delta(t - t')
\; ,
\label{noise-1}
\end{eqnarray}
where the angular brackets denote an average over the random force
realizations. The spectral function for $D_{ij}(\vec x)$ will be specified
below.

We assume that the fluid is incompressible so that the density field
is constant ($\rho ({\vec x},t)=\rho_0$) and $\vec \nabla \cdot {\bf
u}=0$. Under this condition we need only consider the equation for the
conservation of momentum~(\ref{moment1}) and write
\begin{eqnarray}
\frac{\partial {\bf u}}{\partial t} +
 ({\bf u \cdot}\vec \nabla)
{\bf u}
&=& -\frac{1}{\rho_0}\vec \nabla \left[
 p  -\frac{\rho_0}{2} ({\bf \Omega} \wedge {\bf x})^2
\right]
+ \nu \nabla^2 {\bf u}
-
2{\bf \Omega \wedge u}
+ {\bf  f}
\; .
\label{u-field1}
\end{eqnarray}
Notice that the force per unit mass ${\bf f}$ will be taken solenoidal
as well, that is $\vec \nabla \cdot {\bf f}=0$, in order to avoid
having a random component in the pressure.

In the absence of random stirring particular solutions of
Eq.~(\ref{u-field1}) are well known (for an incompressible fluid): it
admits plane wave solutions, called {\it inertial
waves}~\cite{Greenspan,Chandra,LL2}. These are exact solutions of the
non-linear equations, but superposition does not hold. They may have a
role in the transition to turbulence~\cite{Dubrulle}.  In
Subsection~\ref{sec:waves} we will study how the perturbative
corrections modify the propagation of inertial waves.

We now proceed to eliminate the gradient term of Eq.~(\ref{u-field1})
by making use of the incompressibility condition~\cite{McComb}.  We
define the generalized pressure as $p^* \equiv p -\frac{\rho_0}{2}
({\bf \Omega} \wedge {\bf x})^2 $. By taking the divergence of the
previous equation we can solve for $p^*$ to obtain
\begin{equation}
p^* = -\rho_0 \frac{1}{\nabla^2}
\left[
\partial_i \left( u_j \partial_j u_i \right)
+ 2\epsilon_{ijk} \Omega_j \partial_i u_k
\right]
\; ,
\label{p*}
\end{equation}
so that the pressure $p^*$ can be eliminated from Eq.~(\ref{u-field1})
by writing
\begin{equation}
-\frac{1}{\rho_0}\vec \nabla p^* =
\vec \nabla
\frac{1}{\nabla^2}
\vec \nabla \cdot
\left[
 ({\bf u \cdot}\vec \nabla)
{\bf u} + 2 {\bf \Omega \wedge u}
\right]
\; .
\label{grad-p*}
\end{equation}
We can write the Navier-Stokes equation as follows:
\begin{eqnarray}
\frac{\partial \bf u}{\partial t} &+&
\lambda {\cal P}
[
({\bf u \cdot \vec \nabla})
{\bf u}] = \nu \nabla^2 {\bf u} -
{\cal P}(2{\bf \Omega \wedge u}) + {\bf f}
\; ,
\label{u-field2}
\end{eqnarray}
where, following standard practice, we have introduced the
constant $\lambda$ in front of the advective term for book-keeping
purposes~\cite{FNS} ($\lambda$ will be useful when carrying out
the perturbation expansion and is to be set to one
afterwards). The projection operator ${\cal P}$ is given by
\begin{eqnarray}
{\cal P} =  {\bf 1} - {\vec \nabla} \frac{1}{\nabla^2}{\vec
\nabla} \cdot  \; , \label{project}
\end{eqnarray}
and ensures that the non-linear and Coriolis terms are solenoidal.
In Eq.~(\ref{u-field2}), if ${\bf u}$ is solenoidal so is ${\bf
f}$ and vice versa.

Unlike Eqs.~(\ref{moment1}) or~(\ref{u-field1}), Eq.~(\ref{u-field2})
is translation invariant.  That is, the centrifugal term
in~(\ref{moment1}) clearly distinguishes the origin $({\bf x = 0})$ as
a special point; but as we have seen in Eq.~(\ref{grad-p*}) we can
include this term into the generalized pressure and eliminate $p^*$
from the equation. This yields Eq.~(\ref{u-field2}) in which a
preferred {\it direction} (but no preferred point) is singled out by
the angular velocity. This latter equation is invariant under
translations, hence, we can make use of the Fourier transform. Since
in Fourier space
\begin{eqnarray}
{\cal P}_{ij}({\bf k}) &=&   \delta_{ij} - \frac{k_i k_j}{k^2} \;
, \label{project-fourier}
\end{eqnarray}
Eq.~(\ref{u-field2}) only contains vectors orthogonal to ${\bf k}$
and we may refer to this equation as the {\it transverse}
Navier-Stokes equation.

We choose the random force spectrum~(\ref{noise-1}) as follows:
\begin{eqnarray}
\langle f_i (\vec k,\omega) \rangle =
0
\;
\; \; \; \; \; \;
{\rm and}
\; \; \; \; \; \;
\langle f_i(\vec k,\omega)f_j(\vec k',\omega') \rangle =
(2D) k^{-y}(2\pi)^{d+1}{\cal P}_{ij}({\bf  k})\, \delta(\omega +
\omega') \, \delta^{d}(\vec k + \vec k')
\; ,
\label{unoise}
\end{eqnarray}
where $k=\vert \vec k \vert$, $D > 0$ is a measure of the amplitude of
the random force and the real exponent $y>-2$ characterizes the random
force spectrum~\cite{YO}. When $y=d$ the velocity correlations produce
an energy spectrum resembling the Kolmogorov spectrum.  The random
force acts {\it isotropically} on the fluid, thus, whatever
anisotropies emerge at large scale must be due to the Coriolis
force. In the following Sections we analyse the nature of the
anisotropies by solving Eq.~(\ref{u-field2}) in perturbation theory.

\subsection{The Fourier transformed equation}
\label{sec:fourier}

In what follows we use the Fourier transformed equation of
motion~(\ref{u-field2}). Our convention for the Fourier transform is
given by
\begin{equation}
u_j(\vec x,t) = \int_{k < \Lambda}
\frac{d^d \vec k}{(2\pi)^d}\, \int_{-\infty}^{\infty}
\frac{d\omega}{2\pi}\,
u_j(\vec k,\omega)\,
e^{i(\vec k \cdot \vec x - \omega t)}
\; .
\label{fourier}
\end{equation}
We have introduced a wave-number cut-off $\Lambda$, so that the
integral over $\vec k$ is restricted to the values $\vert \vec
k\vert < \Lambda$. The inverse of this cut-off, $1/\Lambda$, can
be associated with the dissipation (Kolmogorov) scale, and
we assume that $1/\Lambda \ll L$~\cite{YO}.

In order to write the Navier-Stokes Eq.~(\ref{u-field2}) in
wave-number representation we transform the ${\bf u}$ field according
to~(\ref{fourier}), apply convolution to the non-linear term, and
invert the Fourier transform to obtain
\begin{eqnarray}
(-i\omega + \nu k^2)u_i(\vec k,\omega) + {\cal P}_{ij} ({\bf k})
(2{\bf \Omega \wedge u})_j &=&
-\frac{i}{2}\lambda [{\cal P}_{ik} ({\bf k}) k_j
+ {\cal P}_{ij} ({\bf k}) k_k]
\int_{p < \Lambda}
\frac{d^d \vec p}{(2\pi)^d}\, \int_{-\infty}^{\infty}
\frac{d\omega'}{2\pi}\,
u_j(\vec k - \vec p,\omega - \omega')
u_k(\vec p,\omega')
\nonumber \\
&+&
f_i(\vec k,\omega)
\; .
\label{fourier-u2}
\end{eqnarray}
This equation, but without the Coriolis term, is a familiar expression
in turbulence research~\cite{FNS,YO,McComb}.  Eq.~(\ref{fourier-u2}) can
be iterated to any desired order in $\lambda$ and will serve as the
starting point for constructing the perturbation expansion, which is
considered in the following Subsection.

\subsection{The linear response function}
\label{sec:response}

If we ``shut-off'' the non-linear terms (proportional to $\lambda$)
in~(\ref{fourier-u2}) we can identify the (inverse) linear response
function, which will be used in carrying out the perturbative
calculation. Thus, setting $\lambda=0$ in Eq.~(\ref{fourier-u2}) we
have
\begin{eqnarray}
(-i\omega + \nu k^2)u_i^{(0)}
(\vec k,\omega) + {\cal P}_{ij} ({\bf k})
(2{\bf \Omega \wedge u}^{(0)})_j &=&
f_i(\vec k,\omega)
\; ,
\label{fourier-u-free}
\end{eqnarray}
where ${\bf u}^{(0)}$ is the linear velocity field. The previous
equation can be written as
\begin{equation}
[G_0(\vec k,\omega)]^{-1}{\bf u}^{(0)}(\vec k,\omega) = {\bf f}(\vec k,
\omega)
\; ,
\label{g-inv1}
\end{equation}
which defines the linear (inverse) response function matrix,
and allows one to solve for the linear velocity field ${\bf
u}^{(0)} (\vec k,\omega)$.  In the next Subsection we calculate
the perturbative corrections to the linear inverse response
function due to the existence of non-linearities and random
forcing, and establish their effect on the physical parameters
({\it e.g.,} viscosity) describing the transverse Navier-Stokes
equation.

\subsection{The non-linear response function and its
first order perturbative expansion}
\label{sec:perturbative}

\begin{figure}
\centerline{\epsfbox{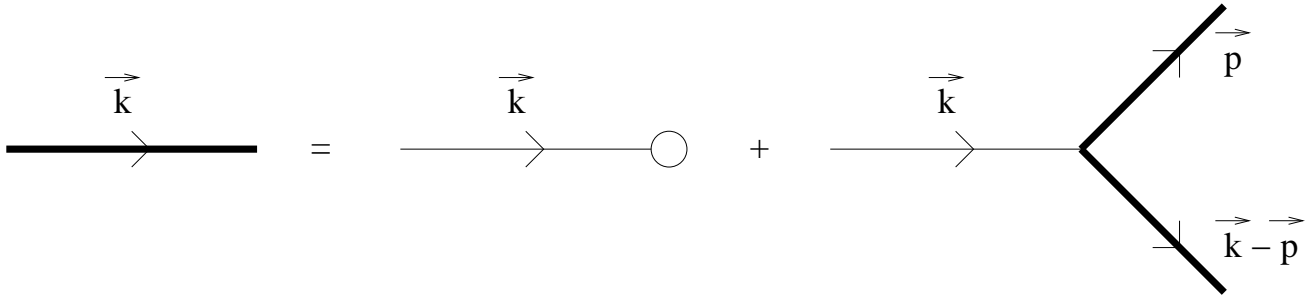}}
\caption{Exact hydrodynamic equation.}
\label{fig1}
\end{figure}

In this Subsection we proceed to expand Eq.~(\ref{fourier-u2}) by
iteration, in powers of the non-linear coupling (see the diagrammatic
representation, Fig.~\ref{fig1}). In this way one can calculate the
perturbative corrections to the response function to any desired order
in $\lambda$. We do not present all the details of such expansion as
these can be found in review articles~\cite{SmithWood} and
textbooks~\cite{McComb}.

\begin{figure}
\centerline{\epsfbox{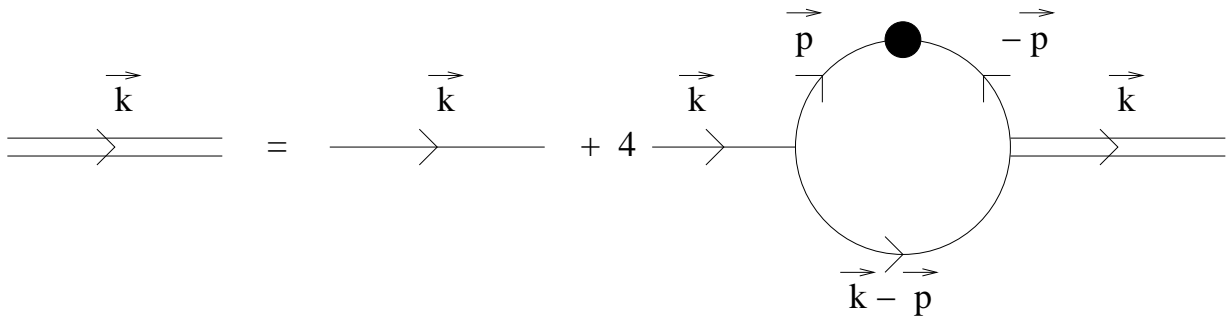}}
\caption{Recursion relation for the first order response function.}
\label{fig2}
\end{figure}

The calculation of the non-linear response function, $[G]$,
requires the correction matrix, $[M]$, which is defined by a
recursion formula:
\begin{equation}
[G] \equiv [G_0] + [G_0] [M] [G] = [G_0] + [G_0] [M] [G_0]
+ [G_0] [M] [G_0] [M] [G_0] + \cdots \; ,
\label{gg}
\end{equation}
which implies
\begin{equation}
{[G]}^{-1} = [G_0]^{-1} - [M]
\label{inverse}
\; .
\end{equation}
In Appendix~\ref{sec:app-diag} we present the diagrammatic
representation of the exact transverse Navier-Stokes equation, as
well as the first order correction to the response function. The
diagram of Eq.~(\ref{gg}) with the correction $[M]$, calculated to
first order of perturbation theory in the random force amplitude,
is presented in Fig.~\ref{fig2}. When written in components, the
correction matrix $[M]$ is given by
\begin{equation}
M_{mn} (\vec k,\omega) =  \left(\frac{-i \lambda}{2} \right)^2 \times 4
\times
[k_r {\cal P}_{mj}({\bf k}) + k_j {\cal P}_{mr}({\bf k})]\,
I_{rjn}(\vec k, \omega)
\; ,
\label{m}
\end{equation}
where the factor $4$ is combinatorial (see Fig.~\ref{fig2}), and
the function $I_{rjn}(\vec k, \omega)$ is defined by
\begin{eqnarray}
I_{rjn}(\vec k, \omega)= \int_{1/L < \vert \vec p  \vert <
\Lambda} \frac{d^d \vec p}{(2\pi)^d}\, \int_{-\infty}^{\infty}
\frac{d\omega'}{2\pi} && [(k_s-p_s) {\cal P}_{ln}({\bf k -\bf p})
+ (k_n -p_n) {\cal P}_{ls}({\bf k - \bf p})] \nonumber
\\
&&
\times
(2D) \, \vert \vec p \vert^{-y}\,
[G_0(\vec p,\omega')]_{ra}{\cal P}_{ab}({\bf p})
[G_0 (-\vec p,-\omega')]_{bs}^T\,
[G_0(\vec k - \vec p, \omega-\omega')]_{jl}
\; ,
\label{loopint}
\end{eqnarray}
(where $T$ means the transposed matrix). The integral over $\vec p$
would be divergent at $\vec p = \vec 0$, so it is cut off at the low
wave-number given by the inverse of the system size scale, $1/L$. The
evaluation of Eq.~(\ref{loopint}) also involves an integration over
the frequency $\omega'$, which has been carried out for {\it
slow rotation} (see Appendix~\ref{sec:app-perturbation}).

As we are interested in the late time and large scale limits we have
computed $M_{mn} (\vec k , \omega)$ only for $\omega=0$ and up to
second order in $\vec k$~\cite{FNS}.  The integration over wave-number
$\vec p$ can be decomposed into an integration over modulus and
angles. The integrand must be expanded, up to second order in $\vec
k$, as this is sufficient to obtain the effective viscosity [see
Eq.~(\ref{diagonal})].  The calculation of $I_{rjn}(\vec k, \omega)$
is straightforward but tedious and its technical details are presented
in Appendix~\ref{sec:app-perturbation}.

According to Eq.~(\ref{inverse}), the matrix $[M]$ provides the order
$\lambda^2$ correction to the inverse response function.  To obtain
$[M]$ (for $\omega=0$ and in the limit ${\bf k} \rightarrow {\bf 0}$),
we carry out the integral of Eq.~(\ref{loopint}) and substitute into
Eq.~(\ref{m}):
\begin{mathletters}
\begin{eqnarray}
M_{ij}(\vec k,0) &=& -\lambda^2 \frac{(2D)S_d}{(2 \pi)^d(2 \nu)^2}
\frac{\Lambda^{d-y-4}-(1/L)^{d-y-4}}{(d-y-4)d(d+2)}(d^2-y-4)\, k^2
\, {\cal P}_{ij}({\bf k}) \label{one-loop-g-inv-isotropica}
\\
&+& \lambda^2 \frac{(2D)S_d\, (2 \Omega_m)}{(2 \pi)^d(2 \nu)^3}
\frac{\Lambda^{d-y-6}-(1/L)^{d-y-6}}{(d-y-6)d(d+2)} (-d^2+d+2)
[k_nk_j \epsilon_{nmi} -k^2 \epsilon_{nmj} {\cal P}_{in}({\bf k})]
\label{one-loop-g-inv-isotropicb} \; ,
\end{eqnarray}
where $S_d$ is the surface area of the unit sphere in $d$ dimensions,
$\Lambda$ and $1/L$ are, respectively, the upper and lower wave-number
cut-offs, $\lambda = 1$, and $\epsilon_{ijk}$ the Levi-Civita tensor
for $d=3$.  In arriving at this final form we have assumed slow
rotation, that is, we have kept the rotation dependent terms up to
$O(\Omega)$ in the calculation. This is explained in
Appendix~\ref{sec:app-perturbation} and we will come back to this
point in Section~\ref{sec:discussion}. We can compare the linear
inverse response function~(\ref{inverse0}) to its first order
correction $M_{ij}(\vec k,0)$. In order to do so, we set $\omega =0$
in Eq.~(\ref{inverse0}) to obtain
\end{mathletters}
\begin{eqnarray}
[G_0(\vec k,0)]^{-1}_{ij} &=&
 \nu k^2 \, \delta_{ij}
+ 2\epsilon_{nmj} \Omega_m \, {\cal P}_{in}({\bf k})
\; ,
\label{inverse0c-projected}
\end{eqnarray}
which is split into an isotropic part (proportional to $\nu$) and an
anisotropic part (proportional to $\Omega$).  The first part of
$[M]$~(\ref{one-loop-g-inv-isotropica}) corrects the molecular shear
viscosity $\nu$, exactly as in the absence of rotation~\cite{FNS}. The
non-linear inverse response function must take the same form as
$[G_0(\vec k,0)]^{-1}_{ij}$, with $\nu$ replaced by $\nu'$, the
effective viscosity~\cite{FNS}.  The value of $\nu'$ can be obtained
by making use of Eqs.~(\ref{inverse})
and~(\ref{one-loop-g-inv-isotropica}), so that
\begin{eqnarray}
\label{eff-vis-1} \nu' k^2 \, \delta_{ij} &=& \nu k^2 \,
\delta_{ij} + \frac{(2D)S_d}{(2 \pi)^d(2 \nu)^2}
\frac{\Lambda^{d-y-4}-(1/L)^{d-y-4}}{(d-y-4)d(d+2)}(d^2-y-4)\, k^2
\, {\cal P}_{ij}({\bf k}) \; .
\end{eqnarray}
If we multiply Eq.~(\ref{eff-vis-1}) by $ {\cal P}_{mi}({\bf k})$ and
make use of the fact that ${\cal P}$ is idempotent, we obtain
\begin{equation}
\nu' = \nu +  \frac{D}{2 \nu^2} \,
\frac{\Lambda^{d-y-4}-(1/L)^{d-y-4}}{d-y-4}\,
\frac{(d^2-y-4)S_d}{d(d+2)(2\pi)^d} \; , \label{nu'}
\end{equation}
which shows that the isotropic term of $[M]$,
Eq.~(\ref{one-loop-g-inv-isotropica}), renormalizes the molecular
shear viscosity $\nu$.

We must distinguish three cases, namely, $y < d-4$, $y=d-4$, and
$y > d-4$. If $y=d-4$, naive perturbation theory fails.  Given
that $\Lambda L \gg 1$, in the case $y < d-4$, the term
$\Lambda^{d-y-4}$ dominates over $(1/L)^{d-y-4}$, so we could take
the limit $L \rightarrow \infty$ (the $\vec p$ integral is
convergent in the lower limit). In contrast, in the case $y >
d-4$, the term $(1/L)^{d-y-4}$ dominates and we instead neglect
the contribution proportional to $\Lambda^{d-y-4}$. In
consequence, the actual expansion parameter is
$D{\Lambda^{d-y-4}}/\nu^3$ or $D{L^{y+4-d}}/\nu^3$, according to
whether $y<d-4$ or $y > d-4$, respectively. We will take $d-4 < y
< d^2-4$. The number $D{L^{y+4-d}}/\nu^3$ can be identified with
the cube of the Reynolds number, on account of the interpretation
of $L$ as the system size scale, and that the dissipation
rate is proportional to the amplitude of the random force
$D$~\cite{YO}.

In particular, in $d=3$,
\begin{equation}
\nu' = \nu - \frac{D}{2 \nu^2} \, \frac{(1/L)^{-y-1}}{y+1}\,
\frac{(y-5)S_3}{15(2\pi)^3} =\nu \left(1 + \frac{1}{60 \pi^2} \,
\frac{5-y}{y+1} \,\frac{D L^{y+1}}{\nu^3}\right)\; .
\label{nu'-3d}
\end{equation}
The most interesting case is $y=d=3$, which yields the Kolmogorov
energy spectrum.

The second part of the correction~(\ref{one-loop-g-inv-isotropicb})
has the same structure as the anisotropic term on the right-hand side
of Eq.~(\ref{inverse0c-projected}), on considering that the first term
within the square brackets vanishes when it acts on $u_j$.  However,
it does not correspond to a renormalization of ${\bf \Omega}$, owing
to the additional dependence on $k$, namely, the $k^2$ factor.  In
fact, this factor is adequate for a ``viscosity term'' that depends on
${\bf \Omega}$ (anisotropy). Therefore, it suggests the introduction
of an anisotropic ``viscosity'', which we address in the following
Section.

\section{Symmetry requirements: the effective ``viscosity tensor''}
\label{sec:effective-tensor}

In the previous Section we have seen that the
term~(\ref{one-loop-g-inv-isotropicb}) induces no corrections to
$\nu$. In this Section we give a physical interpretation of this
anisotropic contribution, and show that it arises from an effective
``viscosity tensor'' for the rotating fluid with mixed symmetry in its
indices.  For a Newtonian fluid the linear relation between the rate
of strain and stress tensors involves a rank four viscosity tensor, so
that we can write~\cite{Batchelor}
\begin{equation}
T_{ij} = \frac{1}{2} \eta_{ijmn}
\,
\left( \frac{\partial u_m}{\partial x_n}
+
\frac{\partial u_n}{\partial x_m}
\right)
\equiv
 \eta_{ijmn}
u_{mn}
\; .
\label{newton}
\end{equation}
As both $T_{ij}$ and $u_{mn}$ are symmetric tensors the only
symmetries of the ``viscosity tensor'' are the following: $
\eta_{ijmn}= \eta_{jimn}$ and $\eta_{ijmn}= \eta_{ijnm}$.  {From} the
above symmetries we conclude that the ``viscosity tensor'' has $36$
independent components (in $d=3$).  We write $\eta_{ijmn}$ as a sum of
a pair-symmetric (S) and a pair-antisymmetric (A) part as follows
\begin{eqnarray}
\eta_{ijmn}
&=&
\frac{1}{2}(\eta_{ijmn} +\eta_{mnij} )
+
\frac{1}{2}(\eta_{ijmn} - \eta_{mnij} )
\equiv
\eta^{S}_{ijmn} + \eta^{A}_{ijmn}
\; ,
\label{viscosity}
\end{eqnarray}
so that $\eta^{S}_{ijmn}$ has the same symmetries as $\eta_{ijmn}$
plus $\eta^{S}_{ijmn}=\eta^{S}_{mnij}$, and $\eta^{A}_{ijmn}$ has the
same symmetries as $\eta_{ijmn}$ plus
$\eta^{A}_{ijmn}=-\eta^{A}_{mnij}$. There are $21$ pair-symmetric and
$15$ pair-antisymmetric independent components.  The presence of axial
symmetry (induced by the Coriolis term) reduces the number of
independent components of both $\eta^S_{ijmn}$ and
$\eta^A_{ijmn}$. The most general axisymmetric tensor (in $d=3$) can
be constructed from $\Omega_i$, $\delta_{ij}$, and $\epsilon_{ijk}$ as
follows
\begin{mathletters}
\begin{eqnarray}
\eta_{ijmn}^S
&=&
\alpha_1 (\Omega^2) (\delta_{im}\delta_{jn} + \delta_{in}\delta_{jm})
\nonumber
\\
&+&
\alpha_2 (\Omega^2) \delta_{ij}\delta_{mn}
\nonumber
\\
&+&
\alpha_3 (\Omega^2) (\Omega_{i} \Omega_j \delta_{mn}
+\Omega_{m} \Omega_n \delta_{ij})
\nonumber
\\
&+&
\alpha_4 (\Omega^2) (\Omega_{i} \Omega_m \delta_{jn}
+\Omega_{j} \Omega_m \delta_{in}
+
\Omega_{i} \Omega_n \delta_{jm}
+\Omega_{j} \Omega_n \delta_{im})
\nonumber
\\
&+&
\alpha_5 (\Omega^2) \Omega_{i} \Omega_{j}\Omega_{m}\Omega_{n}
\; ,
\label{nus}
\\
\eta_{ijmn}^A
&=&
\beta_1 (\Omega^2)
\Omega_q
(\epsilon_{qim}\delta_{jn} +
\epsilon_{qin}\delta_{jm} +
\epsilon_{qjm}\delta_{in} +
\epsilon_{qjn}\delta_{im} )
\nonumber
\\
&+&
\beta_2 (\Omega^2)
\Omega_q
(\epsilon_{qim}\Omega_{j}  \Omega_{n}
+
\epsilon_{qin} \Omega_{j}  \Omega_{m}  +
\epsilon_{qjm}\Omega_{i}  \Omega_{n}  +
\epsilon_{qjn}\Omega_{i}  \Omega_{m}  )
\nonumber
\\
&+&
\beta_3 (\Omega^2)
(\Omega_{i} \Omega_j \delta_{mn}
-\Omega_{m} \Omega_n \delta_{ij})
\label{nua}
\; .
\end{eqnarray}
We observe that the number of independent components has been reduced
from $21$ to $5$ for the pair-symmetric term and from $15$ to $3$ for
the pair-antisymmetric part. The symmetry arguments used in
deducing~(\ref{nus},\ref{nua}) are analogous to those used in the
theory of elasticity~\cite{LL}.
\end{mathletters}

The coefficient functions $\alpha_i(\Omega^2)$ and
$\beta_i(\Omega^2)$ can be written as a series in $\Omega^2$:
\begin{eqnarray}
\alpha_n= \sum_{r=0}^{+ \infty}
\alpha_{nr} (\Omega^2)^r
\; ,
\; \;  \;
\beta_n= \sum_{r=0}^{+ \infty}
\beta_{nr} (\Omega^2)^r
\; .
\label{beta}
\end{eqnarray}
In the limit of slow rotation (linear order in $\Omega$) these
coefficients reduce to their constant ($\Omega=0$) value and,
furthermore, the terms proportional to $\alpha_3, \alpha_4, \alpha_5,
\beta_2,$ and $\beta_3$ vanish at this order.  We can then write for the
``viscosity tensor''
\begin{mathletters}
\begin{eqnarray}
\eta_{ijmn}^S
&=&
\alpha_1 (\delta_{im}\delta_{jn} + \delta_{in}\delta_{jm})
+ \alpha_2 \delta_{ij}\delta_{mn}
\; ,
\label{nus-1}
\\
\eta_{ijmn}^A
&=&
\beta_1
\Omega_q
(\epsilon_{qim}\delta_{jn} +
\epsilon_{qin}\delta_{jm} +
\epsilon_{qjm}\delta_{in} +
\epsilon_{qjn}\delta_{im} )
\label{nua-1}
\; .
\end{eqnarray}
The pair-symmetric part is isotropic and we can identify
$\rho_0\kappa= \alpha_2 + 2\alpha_1/d$ and $\rho_0\nu=\alpha_1$, where
$\kappa$ and $\nu$ denote the molecular bulk and shear kinematic
viscosities, respectively; the pair-antisymmetric part, proportional
to the coefficient $\beta_1$, is identified below.  Incompressibility
means that $\kappa= 0$, leaving only $\nu$, in the absence of
rotation. This is the molecular viscosity in the original
Navier-Stokes equation~(\ref{moment1}), which gets renormalized,
becoming an effective (rotation independent) viscosity, as seen in the
previous Subsection~\ref{sec:perturbative}.  However, it is to be
expected that perturbation theory, at sufficiently high order,
generates a dependence of $\alpha_1$ on $\Omega$, such that one would
be led to consider a rotation dependent viscosity. At the order we are
working, the effect of rotation is to generate a rotation dependent
anisotropic ``viscosity'' of the type~(\ref{nua-1}), as will be
demonstrated below.
\end{mathletters}

The anisotropic ``viscosity''~(\ref{nua-1}) is associated to a stress
tensor $T^A_{ij}$, which yields the following viscous force
\begin{eqnarray}
F_i \equiv \partial_j T^{A}_{ij}&=& \partial_j \left(
 \eta^{A}_{ijmn}
u_{mn} \right)
\, .
\label{ancorr}
\end{eqnarray}
In wave-number representation this can be written as
\begin{eqnarray}
F_i = -\left( \frac{1}{2} \right)
k_j \eta^{A}_{ijmn} (k_n \delta_{mq} + k_m \delta_{nq})\, u_q
= - k_j k_m\eta^{A}_{ijqm}\, u_q
\; ,
\label{ufield1}
\end{eqnarray}
where we have made use of the symmetry within pairs of indices of the
tensor $\eta^A$.  In order to be consistent with the transverse
Navier-Stokes equation, we must project Eq.~(\ref{ufield1}) by means
of ${\cal P}({\bf k})$
\begin{eqnarray}
{\cal P}_{ip} ({\bf k})
\; F_p =
- k_j k_m {\cal P}_{ip}({\bf k}) \eta^{A}_{pjqm}\, u_q
\; .
\label{proj-u-field1}
\end{eqnarray}
We now substitute Eq.~(\ref{nua-1}) into the previous expression to
obtain
\begin{eqnarray}
{\cal P}_{ip}  ({\bf k}) \;  F_p
&=&
- \beta_1
{\cal P}_{ip}({\bf k})\,
k_j \, k_m
 \Omega_n
\big(\epsilon_{pqn} \delta_{jm}+
\epsilon_{pmn} \delta_{jq}+
\epsilon_{jqn} \delta_{pm}+
\epsilon_{jmn} \delta_{pq}
\big)\, u_q
\nonumber \\
&=& -{\beta_1}\,
\Omega_n \,
[k_q k_m \epsilon_{i mn}
+k^2 \epsilon_{mqn} {\cal P}_{im}({\bf k})] \, u_q
= -{\beta_1}\,
\Omega_n \,
k^2 \epsilon_{mqn} {\cal P}_{im}({\bf k}) \, u_q
\; ,
\label{proj-u-field2}
\end{eqnarray}
where we have taken into account the antisymmetric properties of the
Levi-Civita tensor and the fact that ${\cal P}_{ij}({\bf k})k_i =
{\cal P}_{ij}({\bf k})k_j=0$.  We now compare this equation with the
first order perturbative correction of the linear inverse response
function. The force given by Eq.~(\ref{proj-u-field2}) agrees
identically with the force provided by the anisotropic part of the
correction $M_{ij}(\vec k, 0)$, namely, the product of its expression
in~(\ref{one-loop-g-inv-isotropicb}) and $u_j$, once we choose the
coefficient in~(\ref{nua-1}) to be
\begin{equation}
{\beta_1} \equiv -\rho_0\, \frac{(2D) (2 S_d)}{(2 \pi)^d(2 \nu)^3}
(-d^2+d+2) \frac{\Lambda^{d-y-6}-(1/L)^{d-y-6}}{(d-y-6)d(d+2)} \;
. \label{coeff}
\end{equation}
We have thus shown that the
contribution~(\ref{one-loop-g-inv-isotropicb}) to the response
function can be understood as arising from the anisotropic
``viscosity'' tensor~(\ref{nua-1}). Since we are considering the
case $y > d-4$, we can neglect the term $\Lambda^{d-y-6}$. The
expansion parameter, proportional to $D/\nu^3$, must again be
identified with the cube of the Reynolds number. We postpone this
identification to the following Section.

If we set $d=3$ in the expression for $\beta_1$, we obtain
\begin{equation}
{\beta_1} = \rho_0\, \frac{(2D) (2 S_3)}{(2 \pi)^3(2 \nu)^3} 4
\frac{(1/L)^{-y-3}}{(y+3)15} =  \frac{\rho_0}{15 \pi^2(y+3)}\,
\frac{D L^{y+3}}{\nu^3} \; . \label{coeff-3d}
\end{equation}

\section{Effective large scale dynamics}
\label{sec:effective}

In this Section we focus on the new force $F_i = \partial_j T_{ij}^A$
that appears in the fluid equation of motion due to first order
perturbative corrections, and discuss some possible physical
implications.

\subsection{The quasi-local force}
\label{sec:axial-force}

Here we use the results derived in previous Sections to
``correct'' the transverse Navier-Stokes equation with the newly
generated (first order in $\Omega$) terms that arise at large scales
due to rotation. In coordinate space the force~(\ref{ancorr}) becomes
\begin{eqnarray}
F_i &=&
{\beta_1}
\left[
-(\vec \Omega \wedge \nabla^2 \vec u)_i
+  \partial_i [\vec \Omega \cdot (\vec \nabla \wedge \vec u)]
\right]
\; .
\label{pseudo}
\end{eqnarray}
In deriving this equation we have made use of the incompressibility
condition and the antisymmetry properties of the Levi-Civita
tensor. This is a quasi-local~\footnote{Quasi-local, in general, means
depending on a function (in this case the velocity field ${\bf u}$)
and its derivatives.} force (per unit volume) in which the angular
velocity couples to the Laplacian of the fluid velocity and to the
vorticity ${\bm\omega} = \vec \nabla \wedge {\bf u}$.

The physical character of the force $F_i$ can be revealed by writing
the {\it effective} equation of motion in coordinate space. We have
\begin{eqnarray}
\frac{\partial \bf u}{\partial t} &+& \lambda {\cal P}({\bf u\cdot
\vec \nabla u}) = \nu' \nabla^2 {\bf u} - {\cal P}(2{\bf \Omega \wedge
u}) - {\beta'_1}\,{\cal P}({ 2\bf \Omega \wedge \nabla^2 u}) + {\bf f}
\; ,
\label{newmodelu}
\end{eqnarray}
where $\nu'$ is the rotation independent effective kinematic
viscosity, [see Eq.~(\ref{nu'})], and ${\beta'_1} =
{\beta_1}/(2\rho_0)$. The magnitude of the correction linear in
$\Omega$ (the new force $\vec F$) can be determined by comparing it
with either the Coriolis force or the viscosity correction. We obtain
for the ratio to the Coriolis force: ${\beta'_1} \left| 
(\nabla^2 {\bf u}) \right| / \left| {\bf u}\right| \sim D{L^{y+1}}/\nu^3
(L\Lambda)^2$, assuming that the scale of spatial velocity
fluctuations is $1/\Lambda$, that is, $\left| (\nabla^2 {\bf u})
\right| / \left| {\bf u} \right| \sim \Lambda^{2}$.  Hence, the
expansion parameter can be identified with the cube of the Reynolds
number (as in the previous Section) times $(L\Lambda)^2$. On the other
hand, the ratio between the correction linear in $\Omega$ and the
correction linear in $\nu'$ is $\sim \Omega L^{2}/\nu \sim
Ek^{-1}$. This was to be expected, since $Ek$ measures the relative
strength of the viscosity and Coriolis force.

We conclude this Subsection by providing an important property of the
new force. In general, for a stress tensor associated with a
pair-antisymmetric ``viscosity tensor'', the power is given
by~\cite{LL2}
\begin{eqnarray}
P &\propto& \int d^3 \vec x \; u_{ij} T_{ij}^A
= \int d^3 \vec x \;  u_{ij} \eta^A_{ijmn} u_{mn}
\; ,
\end{eqnarray}
but, as $\eta^A_{ijmn}= -\eta^A_{mnij}\, ,$ we conclude that $P=0$;
therefore, this stress tensor, $T_{ij}^A$, does not lead to
dissipation and is not truly viscous. This implies that the name
``viscosity tensor'' is not appropriate, and we have only introduced
it by analogy with the truly viscous pair-symmetric tensor~\cite{LL2}.

\subsection{Inertial waves}
\label{sec:waves}

As already mentioned, rotating incompressible fluids support wave
solutions that are exact solutions of the non-linear
equations~\cite{Greenspan,Chandra}.  Although we have derived the
quasi-local force from the randomly forced equations (that
describe turbulence), it is of interest to see how this force
affects these wave solutions.  Let us consider wave solutions of
the form $e^{i(\vec k \cdot \vec x + \omega t)}$ as single mode
plane waves of the linearized equation~(\ref{newmodelu})
(the incompressibility condition annihilates the advective term)
but without the forcing term
\begin{eqnarray}
\frac{\partial \bf u}{\partial t} &=&
\nu' \nabla^2 {\bf u} -
{\cal P}({2\bf \Omega \wedge u})
- {\beta'_1} {\cal P}( {2\bf \Omega \wedge \nabla^2 u})
\; .
\label{nsp-II-2}
\end{eqnarray}
We can write this equation in components as follows
\begin{equation}
\big(i\omega + \nu' k^2 \big)u_i=
2(-1+{\beta'_1} k^2){\cal P}_{im}({\bf k})\epsilon_{mqn} \Omega_q u_n
\; .
\label{comps-II-2}
\end{equation}
We are free to choose ${\bf \Omega} = (\Omega_x, 0, \Omega_z)$ and
${\bf k} = (0,0,k)$. This choice implies that $u_z = 0$
(incompressibility condition), so that the waves are transverse.
{From} our choice for ${\bf k}$ the only non-vanishing components of
${\cal P}({\bf k})$ are ${\cal P}_{xx}$ and ${\cal P}_{yy}$, which are
equal to one. We can write the $x$ and $y$ components [the $z$
component of~(\ref{comps-II-2}) is {identically null})] as follows
\begin{mathletters}
\begin{eqnarray}
\big(i\omega + \nu' k^2 \big)u_x
&=&
-2(-1 + \beta'_1 k^2)
\Omega_z u_y
\; ,
\label{uu-x-2}
\\
\big(i\omega + \nu' k^2 \big)u_y
&=&
2(-1 + \beta'_1 k^2)
\Omega_z u_x
\; .
\label{uu-y-2}
\end{eqnarray}
These equations yield the (complex) frequency of the plane waves
(dispersion relation)
\end{mathletters}
\begin{equation}
\omega (\vec k) = \pm 2 \Omega \cos \theta (1 - {\beta'_1}k^2) +
i\nu' k^2
\; ,
\label{disp-II}
\end{equation}
where $\theta$ is the angle between ${\bf k}$ and ${\bf \Omega}$, so
that ${\bf k} \cdot {\bf \Omega}=k\Omega \cos \theta$, with $k=\vert
\vec k \vert$ and $\Omega=\vert \vec \Omega \vert$.  By making use of
the dispersion relation~(\ref{disp-II}) it is easy to see that the
waves are circularly polarized and transverse
\begin{equation}
u_y = \pm i u_x \; ,   \qquad u_z = 0
\; .
\end{equation}
Wave packets are not solutions of the non-linear equations and can
only be considered for small wave amplitudes.  The group velocity
${\bf V}$ of a wave packet is (for vanishing viscosity)
\begin{eqnarray}
V_i (\vec k) &\equiv& \frac{\partial \omega (\vec k)}{\partial k_i}
= \pm\frac{2(1-{\beta'_1}k^2)}{k} {\cal P}_{ij}({\bf k}) \Omega_j
\mp 4 \beta'_1
\frac{k_i}{k}
{\bf \Omega \cdot k}
\; ,
\label{group}
\end{eqnarray}
and the phase velocity ${\bf v}$ (for vanishing viscosity)
\begin{eqnarray}
v_i (\vec k) &\equiv& \frac{\omega (\vec k)}{k^2}  k_i
= \pm{2(1-{\beta'_1}k^2)} \frac{\bf \Omega \cdot k}{k^3}
{k_i}
\; .
\label{phase}
\end{eqnarray}
The ``standard textbook'' result is recovered by taking the limit
${\beta'_1} \rightarrow 0$~\cite{Greenspan,Chandra}.  {From} this
calculation we see that the new term in Eq.~(\ref{nsp-II-2}) not only
changes the wave frequency and phase velocity, but also the group
velocity.  Moreover, the group velocity is no longer perpendicular to
the phase velocity
\begin{equation}
{\bf V \cdot k} = \mp 4{\beta'_1} k\, {\bf \Omega \cdot k} \neq 0
\; .
\end{equation}
The quasi-local force {would} cause the energy transport to not be
perpendicular to the phase velocity and a small fraction of the energy
in the waves to be transported parallel to the wave vector.

\section{Discussion}
\label{sec:discussion}

We have applied perturbation theory to a homogeneous
incompressible viscous fluid subject to solid body rotation and
isotropic random forcing.  At small scales, we assume that only
the molecular shear viscosity and the Coriolis force are needed in
writing down the {\it transverse} Navier-Stokes
equation~(\ref{u-field2}). Our (first order) perturbative results
demonstrate that (i) anisotropic components of the effective
``viscosity tensor'' are dynamically generated at large scales by
the combined interplay of the Coriolis force, the random forcing
term, and the inherent non-linearity of the Navier-Stokes
equation, (ii) the molecular shear viscosity $\nu$ gets corrected
in the same manner as for isotropic randomly stirred
turbulence~\cite{FNS,YO,McComb}.

These perturbative results are corroborated by a symmetry principle.
By making use of the axial symmetry of a rotating fluid, we have
constructed the most general ``viscosity tensor'' that is invariant
under such symmetry.  The preferred direction singled out by the
angular velocity ${\bf \Omega}$ breaks the isotropy and leads to new
terms in the ``viscosity tensor'' absent in the isotropic
case~\cite{LL2}. We have also determined and described the role of the
first order perturbative new term in the effective fluid equation of
motion.  It acts as a quasi-local force and, like the Coriolis force,
is not dissipative and does no net work on the fluid.  Most
importantly, we find that this quasi-local force affects the
propagation of inertial waves in rotating fluids. For small wave
amplitudes a fraction of the wave energy is transported in the same
direction as the phase velocity.

The perturbative calculation is developed as a double expansion:
in both the amplitude of the random force and the angular
velocity. The actual dimensionless expansion parameters turn out
to be the cube of the Reynolds number and this number times the
inverse of the Ekman number, respectively.  For simplicity, we
have restricted ourselves to the computation of the lowest order
in both: that is, first order in the random force amplitude and
linear order in ${\bf \Omega}$. This is sufficient to generate
two, out of the total of eight tensor terms (seven, on account of
incompressibility) allowed by axial symmetry [see Eqs.~(\ref{nus})
and~(\ref{nua})]. We conjecture that, at first order in the random
force amplitude, all the remaining tensor terms are generated for
higher powers in $\Omega$: $\alpha_3$, $\alpha_4$, and $\beta_3$
at quadratic order, $\beta_2$ at cubic order, and finally
$\alpha_5$ at quartic order. Of course, as increasing powers of
$\Omega$ are taken into account, the coefficient
functions~(\ref{beta}) must be expanded out to the order of
$\Omega$ being investigated. The higher order terms will be needed
to study the effects of fast rotation and to track the onset of
the bi-dimensionalization of the fluid~\cite{Dubrulle}.

An important step in this direction will be provided by a complete
renormalization group (RG) analysis of the large scales properties of
a rotating incompressible fluid.  In order to carry out this RG
analysis one may need to calculate the (perturbative) corrections to
the non-linear coupling term $\lambda$, as well as the random force
amplitude $D$, and combine these two with the response function
calculation presented in this paper. Once in hand, the RG fixed points
can be determined and the corresponding asymptotic behavior of the
rotating fluid deduced, allowing us to compute quantities such as the
scale-dependent Reynolds and Ekman numbers, among others. We hope to
report on these developments elsewhere.

\section*{Acknowledgments}

The authors thank Juan P\'erez-Mercader and Alvaro Dom\'{\i}nguez
for discussions and for reading an earlier version of this paper.
One of us (DH) acknowledges correspondence with Arjun Berera
covering a wide range of topics in turbulence theory.  
Our work is supported by grant BFM2002-01014 and the work of
Jos\'e Gaite is further supported by a Ram\'on y Cajal contract, both of the
Ministerio de Ciencia y Tecnolog\'{\i}a.

\appendix

\section{Diagrammatic Representation of the Exact Equation and the
first order correction to the response function}
\label{sec:app-diag}

In this Section we include a diagrammatic representation of the exact
transverse Navier-Stokes equation [see Eq.~(\ref{fourier-u2})] and of
the recursion relation for the first order response function [see
Eq.~(\ref{gg})]. We have followed the convention introduced in
Ref.~\cite{FNS}.

The exact equation can be represented as
\noindent
where the thick lines represent ${\bf u}(\vec k, \omega)$, the open
circles represent $f(\vec k, \omega)$, and the thin lines represent
$G_0(\vec k, \omega)$. This graphic representation implies that the
linear velocity field ${\bf u}^{(0)} (\vec k, \omega)$ is given by a
thin line attached to an open circle, as we have ${\bf u}^{(0)} (\vec
k,\omega)= G_0(\vec k,\omega) \, {\bf f} (\vec k, \omega)$.

In this diagrammatic representation we can also depict the recursion
relation for the response function $G_{ij}(\vec k, \omega)$,
Eq.~(\ref{gg}), given by the first order correction $[M]$,
[Eq.~(\ref{m})], as follows
\noindent
where the solid circle represents the random force average $\langle
f_m(\vec p, \omega') f_n(-\vec p, -\omega') \rangle$ introduced in
Eq.~(\ref{unoise}), and the double thin lines represent the first
order response function $[G]$.

\section{ Perturbation expansion and evaluation of the integrals}
\label{sec:app-perturbation}

The large distance and long time renormalizability of the transverse
Navier-Stokes equation~(\ref{u-field2}) implies that the corrected
response function $G$ must have the same mathematical structure as its
linear counterpart $G_0$. $G$ has, therefore, the same frequency and
wave-number dependence as $G_0$ and contains the same number of
parameters.  The renormalization of the response function yields the
(rotation independent) effective viscosity $\nu'$. In this Appendix we
outline the major points in the calculation of the first order
correction to the response function~\cite{FNS}.

Eq.~(\ref{g-inv1}) defines the linear inverse response function, which
is given by
\begin{equation}
[G_0(\vec k,\omega)]^{-1}=
\left( \matrix{ a-b & h & 0 \cr
                f & a+b & 0 \cr
                c & d &  a }
\right)
\; ,
\label{inverse0}
\end{equation}
where the matrix entries are: $a = -i\omega + \nu k^2$, $b = 2\Omega
{k_1k_2}/{k^2}$, $c = -2\Omega {k_2k_3}/{k^2}$, $d = 2\Omega
{k_1k_3}/{k^2}$, $f = 2\Omega (1 - {k^2_2}/{k^2})$, and $h = -2\Omega
(1 - {k^2_1}/{k^2})$. In the limit of vanishing rotation ($\Omega =
0$) we recover an isotropic diagonal matrix for the linear response
function and its inverse~\cite{FNS}
\begin{equation}
[G_0(\vec k,\omega)]^{-1}_{ij} = (-i \omega + \nu k^2) \, \delta_{ij}
\; , \; \; \;
[G_0(\vec k,\omega)]_{ij} = (-i \omega + \nu
k^2)^{-1} \, \delta_{ij}
\; .
\label{diagonal}
\end{equation}
Given the matrix form of the inverse linear response
function~(\ref{inverse0}) we can write for the linear response
function~\footnote{This is also sometimes called the linear
propagator~\cite{FNS,McComb}.}
\begin{equation}
[G_0 (\vec k,\omega)] =
\frac{1}{a^3 - ab^2 -afh}\,
\left( \matrix{ a^2 + ab& -ah & 0 \cr
                -af & a^2-ab & 0 \cr
                -ac-bc+df& -ad+bd+ch& a^2 - b^2 -fh}
\right)
\; .
\label{freeuresp}
\end{equation}
The first order correction to $G$ is given in
Subsection~\ref{sec:perturbative} [see Eqs.~(\ref{gg}),
(\ref{inverse}), (\ref{m}), and~(\ref{loopint})]. We have restricted
ourselves to the limit of slow rotation (linear order in $\Omega$),
and therefore need to expand both $G_0$ and $G_0^{-1}$ to this order.
We point out that $G_0^{-1}$ is already linear in $\Omega$ [see
Eq.~(\ref{inverse0})]. To linear order in $\Omega$ the linear response
function is given by
\begin{equation}
[G_0 (\vec k,\omega)] =
\frac{1}{a^2}\,
\left( \matrix{ a+b& -h & 0 \cr
                -f & a-b & 0 \cr
                -c& -d& a}
\right)
+ O(\Omega^2)
=
\frac{1}{a}\,
\left( \matrix{ 1& 0 & 0 \cr
                0 & 1 & 0 \cr
                0& 0& 1}
\right)
+
\frac{1}{a^2}\,
\left( \matrix{ b& -h & 0 \cr
                -f & -b & 0 \cr
                -c& -d& 0}
\right)
+ O(\Omega^2)
\; .
\label{linear-omega-go}
\end{equation}
The first step of this calculation is to carry out the frequency
integration (over $\omega'$) in Eq.~(\ref{loopint}).  {From}
Eq.~(\ref{inverse0c-projected}) we see that the viscosity $\nu$ is the
coefficient of the $k^2$ term, and to obtain the effective viscosity
we can set the frequency $\omega$ to zero from the outset.  Once we
make $\omega=0$ the integral over $\omega'$ can be computed by means
of the calculus of residues~\cite{FNS}.  We close the contour in the
lower half plane, where there are in general three simple poles, which
coalesce into one double pole in the limit of slow rotation.  We
obtain (keeping up to linear terms in $\Omega$):
\begin{eqnarray}
\label{freq}
\int^{\infty}_{-\infty} \frac{d\omega'}{2\pi}\,
[G_0(\vec p,\omega')]_{rt}[G_0(-\vec p,-\omega')]_{sq}
[G_0(\vec k - \vec p,-\omega')]_{jl} &=& \delta_{rt}\,\delta_{sq}\,
\delta_{jl}
\frac{(1 + \frac{\vec p \cdot \vec k}{p^2})}{(2\nu p^2)^2}
+ \delta_{rt}\,\delta_{sq}
\frac{(1 + \frac{2\vec p \cdot \vec k}{p^2})}{(2\nu p^2)^3}[\Omega(\vec k
-\vec p)]_{jl}
\nonumber \\
&+&
\delta_{rt}\,\delta_{jl}
\frac{(1 + \frac{\vec p \cdot \vec k}{p^2})}{(2\nu p^2)^3}[\Omega(-\vec p)]
_{sq}
+
\delta_{sq}\,\delta_{jl}
\frac{(2 + \frac{3\vec p \cdot \vec k}{p^2})}{(2\nu p^2)^3}[\Omega(\vec p)]
_{rt}
\nonumber \\
&+& O(k^2)
\; .
\label{between-step}
\end{eqnarray}
We have carried out an expansion in $k=\vert \vec k \vert$ and only
kept up to linear order, as Eq.~(\ref{m}) is already linear in $k$,
and we only need to compute the $k^2$ term. The previous equation was
evaluated assuming a positive molecular shear viscosity coefficient
($\nu > 0$). A change in sign will change the location of the poles
and modify the frequency integration.  We have introduced the matrix
$[\Omega (\vec p)]$ defined as follows
\begin{equation}
[\Omega(\vec p)]_{mk} = -2\epsilon_{ijk} \Omega_j \, {\cal P}_{im}({\bf p})
\; ,
\end{equation}
which is already linear in the angular velocity.

If we make use of this intermediate result~(\ref{between-step})
and substitute it into Eq.~(\ref{loopint}), we are left with the
integration over the wave-number $\vec p$. {From}
{Eqs.~(\ref{inverse}) and~(\ref{inverse0c-projected})} we can see
that the renormalization of the molecular viscosity $\nu$ requires
that we expand the first order correction $[M]$ up to second order
in the wave-number $\vec k$.  The factor in square brackets in
Eq.~(\ref{m}) is already linear in $k$, so that the
integral~(\ref{loopint}) only needs to be expanded to first order
in $k$.  It is important to note that this integral depends on
$\vec k$ not only through the integrand but also through its
limits of integration.This is because all wave-numbers appearing
in~(\ref{loopint}) must remain within the set of wave-numbers to
be integrated over (in this case we must ensure that both $\vec p$
and $\vec k - \vec p$ belong to this set)~\cite{FNS}. This means
we must integrate over the {\it intersection} of the domains $1/L
\leq |\vec p| \leq \Lambda$ and $1/L \leq |\vec k - \vec p| \leq
\Lambda$, where $1/L$ is the lower cut-off. To first order in
$\vec k$ the second inequality can be written as $1/L + k\cos
\theta < p < \Lambda + k\cos \theta$, where $\theta$ is the angle
between $\vec k$ and $\vec p$ ($\vec k \cdot \vec p=k\,p \cos
\theta$).  There are two cases to consider: (i) if $\cos \theta >
0$ the intersection of the two intervals can be expressed as the
difference of intervals $[1/L,\Lambda] - [1/L,1/L + k\cos \theta]$
and (ii) if $\cos \theta < 0$ the intersection can be written as
$[1/L,\Lambda] - [\Lambda + k\cos \theta , \Lambda]$. This means
that the complete wave-number integration, valid up to $O(k^2)$,
can be written as
\begin{equation}
\label{momshell} \int_{ 1/L <\vert \vec p \vert < \Lambda \; , \;
1/L < \vert \vec k - \vec p \vert < \Lambda } \frac{d^d\vec
p}{(2\pi)^d} = \int d\Omega_d \, \left(\int_{1/L}^{\Lambda} -
\int_{1/L}^{1/L + k \cos \theta} - \int_{\Lambda + k \cos
\theta}^{\Lambda} \right) \frac{dp\; p^{d-1}}{(2\pi)^d} + O(k^2)
\; ,
\end{equation}
where $d\Omega_d$ is the surface element of the unit sphere in $d$
dimensions.

We also present the following projection operator product expansions
that prove to be useful when handling the intermediate steps of the
computation
\begin{eqnarray}\label{projectexpand}
{\cal P}_{a,bc}({\bf k - \bf p})
{\cal P}_{dc}({\bf p}) &=&
(k_a - p_a)\Big( \delta_{bd} - \frac{p_b p_d}{p^2} \Big) -
\frac{p_a p_b}{p^2} \Big( k_d - \frac{\vec p \cdot \vec k\; p_d}{p^2}\Big)
+ O(k^2)
\; ,
\nonumber
\\
{\cal P}_{a,bc}({\bf k - \bf p})
{\cal P}_{ad}({\bf p}) &=& k_a{\cal P}_{bc}(
{\bf p}){\cal P}_{ad}({\bf p}) + O(k^2)
\; ,
\end{eqnarray}
where
\begin{eqnarray}\label{projectexpand-2}
{\cal P}_{a,bc}({\bf k})
&=&
k_b {\cal P}_{ac}({\bf k}) +
k_c {\cal P}_{ab}({\bf k})
\; .
\end{eqnarray}
Finally, various identities needed for the angular integrations are
collected here.  Let $S_d$ represent the surface area of the unit
$d$-sphere and $\hat n_j$ denote a unit vector in the $j$-th
direction. The only angular integrations required are of the following
types
\begin{eqnarray}\label{sphere}
\int d\Omega_d &=&  S_d
\; , \nonumber \\
\int d\Omega_d \, {\hat n}_i {\hat n}_j &=& \frac{S_d}{d} \delta_{ij}
\; ,
\nonumber \\
\int d\Omega_d \, {\hat n}_i {\hat n}_j {\hat n}_n {\hat n}_m &=&
\frac{S_d}{d(d + 2)} \big( \delta_{ij} \delta_{mn} +
\delta_{im} \delta_{jn} + \delta_{in} \delta_{jm} \big)
\; .
\end{eqnarray}
The angular integration of the product of an odd number of unit
vectors over the unit $d$-sphere vanishes identically.  If we make use
of the previous results, we obtain the first order correction as
written
in~(\ref{one-loop-g-inv-isotropica},\ref{one-loop-g-inv-isotropicb}).

\end{document}